\documentclass[12pt,fleqn, a4paper]{article}

\newcommand{\ra}{\rightarrow}
\newcommand{\bra}{\langle} \newcommand{\ket}{\rangle}
\newcommand{\be}{\begin{equation}}
\newcommand{\ee}{\end{equation}}
\newcommand{\bea}{\begin{eqnarray}}
\newcommand{\eea}{\end{eqnarray}}
\newcommand{\eps}{\epsilon}
\newcommand{\E}{\mbox{e}}

\newcommand{\ffi}{\varphi}

\newcommand{\ep}{\qquad {\vrule height 10pt width 8pt depth 0pt}}

\newcommand{\grintl}{[\kern-.18em [}
\newcommand{\grintr}{]\kern-.18em ]}

\newtheorem{lem}{Lemma}[section]
\newtheorem{prop}{Proposition}[section]
\newtheorem{thm}{Theorem}[section]
\newtheorem{cor}{Corollary}[section]
\usepackage{amsfonts}

\def\un{\hbox{$\mit I$\kern-.77em$\mit I$}}
\def\0{\hbox{$\mit I$\kern-.70em$\mit O$}}
\def\R{{\mathbb R}}
\def\T{{\mathbb T}}
\def\Z{{\mathbb Z}}
\def\N{{\mathbb N}}
\def\C{{\mathbb C}}
\def\E{{\mathbb E}}
\def\P{{\mathbb P}}

\begin{document}

\title{Density of States and Thouless Formula for Random Unitary Band Matrices}
\author{Alain Joye \\ Institut Fourier \\
Universit\'e de Grenoble 1, 
BP 74 \\
38402 Saint-Martin d'H\`eres Cedex, France}
\date{ }
\maketitle
\abstract{We study the density of states measure for some class of random 
unitary band matrices and prove a Thouless formula 
relating it to the associated Lyapunov exponent.
This class of random matrices appears in the study of the 
dynamical stability of certain quantum systems and can also be considered 
as a unitary version of the Anderson model. We further determine 
the support of the density of states measure and
provide a condition ensuring it possesses an analytic density.}

\setcounter{equation}{0}
\section{Introduction}

The stability of quantum dynamical systems generated by time periodic hamiltonians
is sometimes characterized by means of the spectral properties of the corresponding unitary
evolution operator over a period, also called monodromy operator, see \cite{be, h1, c2}. 
Unfortunately, even for this relatively simple time-dependence, except for certain specific 
models, e.g. \cite{c1,df, bo}, it is rarely the case that one has enough information 
about the actual monodromy operator so that a complete
spectral analysis can be performed. Therefore, one resorts to different approximation
techniques in some specific regimes to say something about the spectrum. For example,
KAM inspired techniques, see e.g. \cite{be, c0, ds1, ade, dlsv, gy}, or adiabatic related 
approaches, see e.g. \cite{h2, h3, h4,  n1, j, n2}, have been used to tackle this problem.

In case the complexity of the monodromy operator is important enough to forbid
of a complete description of it, one may resort to a statistical modelization. 
It is the case in particular in the study of
the quantum dynamics of electrons confined to a ring threaded by a time dependent
magnetic flux, see e.g. the paper \cite{bb} and references therein. A modelization
of this dynamics by means of an effective random monodromy operator taking into account
the details of the metallic structure of the ring is considered and tested numerically in 
\cite{bb}. We refer the reader to this paper and \cite{bhj} for a more detailed account of 
the construction of the monodromy operator.

Motivated by this approach, the spectral analysis of a class of random and deterministic unitary 
operators, which contains the above  monodromy operator, is performed in \cite{bhj}. 
The main characteristics of these unitaries is that, when expressed as matrices in some basis, 
they display a band structure: more precisely they are five-diagonal.
While the construction of the set of unitaries studied in \cite{bhj} is patterned after 
the above mentionned physical model, we believe it may be useful for a wider class of problems. 
Moreover, in the regime we consider here, this set of unitaries bears strong resemblances 
with the Jacobi matrices related to self-adjoint discrete one-dimensional Schr\"odinger 
operators.

Another motivation in that direction stems from the recent paper \cite{cmv} where certain unitary 
infinite matrices associated to the construction of orthonormal polynomials on the unit circle 
are shown to display the same five-diagonal structure as our set of monodromy operators. 
These matrices are shown in \cite{cmv} to be unitarily equivalent to unitary operators 
introduced almost ten years ago in \cite{gt} for the study of a related trigonometric 
moment problem. Moreover, in the latter paper, some effects of randomness in the 
coefficients of these operators were investigated.\\

The goal of the present paper is to pursue the analysis of such random unitaries 
in the setting considered in the paper \cite{bhj}. 
The {\it phases} of the matrix elements of the 
five-diagonal operators are random variables and the deterministic
modulus depend on one parameter only: if the phases are all set to zero, what 
we will call the "free case", the unitary operator depends on a "reflexion" 
coefficient $r\in ]0,1[$, see below.  However, while the analysis of \cite{bhj} 
focused on spectral issues, i.e. proving singularity of the almost sure spectrum by means 
of a unitary version of the Ishii-Pastur theorem and the positivity of the Lyapunov 
exponent obtained via Furstenberg's Theorem, the main object of the present 
study of the density of states measure and its links with the corresponding 
Lyapunov exponent.

More precisely, expressing the density of states as the density of eigenvalues of 
a series of unitary operators restricted to "boxes",
we are able to state this relation as what is known as a Thouless formula. This 
formula allows to compute the Lyapunov exponent by means of the density of states
and to recover the a.c. component of the density of 
states measure by means of a derivative of the Lyapunov exponent. 
A consequence 
of our version of Thouless formula is the extension of some results of \cite{bhj} 
providing, in particular, an explicit value of the 
Lyapunov exponent in these cases. We also prove the validity of the Thouless
formula for the deterministic free case, by explicit computations of the relevant 
quantities. 
Taking advantage of the analogy of our unitary matrices with the one 
dimensional discrete Schr\"odinger operator, we characterize the support of the
density of states in term of that of the distribution of the i.i.d. random phases. 
Finally, we provide an effective criterion ensuring analyticity of the integrated 
density of states in terms of the exponential decay rate of the Fourier coefficients
of the the distribution of the phases. This result relies on some kind of
propagation estimates for the free evolution.

We note here also that a Thouless formula is proven for the unitary random operator 
studied by Geronimo and
Teplyaev in \cite{gt}. The corresponding random matrix $V_{\omega}$ is 
defined in the canonical basis of $l^2(\Z)$ as well but displays a different
structure: for all $k\in \Z$, the vector $V_{\omega}\ffi_k$ has non zero 
coefficients along $\ffi_j$, for $j=-\infty,\cdots, k+1$ only. Such matrices
are also called Hessenberg matrices.
However, the operator under consideration here presents different characteristics 
from the one of \cite{gt}, or \cite{cmv}, in particular regarding the way randomness 
appears in the phases of the matrix elements.

The plan of the paper is as follows. Section 2 is devoted to the definition of
the model and its basic properties. The density of states is introduced in the next
section and Thouless formula is proven in Section 4. The statements about the
support of the density of state and ist analyticity properties are made in Section
5, whereas an Appendix contains some technical items.

\section{The Model}
We present here the unitary matrices we will be concerned 
with and recall some of its basic properties to be used later.

The unitary operator we consider has the following explicit form
in the canonical basis $\{\ffi_k\}_{k\in\Z}$ of $l^2(\Z)$
\bea\label{matel}
U_{\omega}\ffi_{2k}&=&irt e^{-i\eta_{2k}^{\omega}}\ffi_{2k-1}
+
r^2 e^{-i\eta_{2k}^{\omega}}\ffi_{2k}\nonumber\\
&+&irt e^{-i\eta_{2k+1}^{\omega}}\ffi_{2k+1}
-
t^2 e^{-i\eta_{2k+1}^{\omega}}\ffi_{2k+2}\nonumber\\
& & \nonumber\\
U_{\omega}\ffi_{2k+1}&=&-t^2 e^{-i\eta_{2k}^{\omega}}\ffi_{2k-1}
+
itr e^{-i\eta_{2k}^{\omega}}\ffi_{2k}\nonumber\\
&+&r^2 e^{-i\eta_{2k+1}^{\omega}}\ffi_{2k+1}
+
irt e^{-i\eta_{2k+1}^{\omega}}\ffi_{2k+2},
\eea
for any $k\in\Z$. According to \cite{bhj}, the random phases 
$\{\eta_k^{\omega}\}_{k\in\Z}$ are functions of some physically relevant
i.i.d. random variables $\{(\theta_k^{\omega}, \alpha_k^{\omega})\}_{k\in\Z}$ 
on the torus given by 
\be\label{defeta}
\eta_k^{\omega}=\theta_k^{\omega}+\theta_{k-1}^{\omega}+
\alpha_k^{\omega}-\alpha_{k-1}^{\omega},
\ee
for all $k\in\Z$ and the coefficients $r, t\in]0,1[$ are interpreted as 
reflexion and transition coefficients linked by $r^2+t^2$. 
We will identify the operator and its matrix representation (\ref{matel}). Let 
us recall that these parameters are assumed to be different from their extreme 
values $0$ and $1$, because in case $r=1\Longleftrightarrow t=0$ the operator 
$U_{\omega}$ is diagonal and if $r=0\Longleftrightarrow t=1$, it is unitarily 
equivalent to the direct sum of two shifts. Let us finally mention that 
$U_{\omega}$ is constructed in section 2 of \cite{bhj} as a product of two unitaries 
given by infinite direct sums of $2\times 2$ unitary blocks.

\subsection{Ergodic properties}
More precisely, let us introduce a probabilistic space 
$(\Omega, {\cal F}, \P)$, where $\Omega$ is identified with $\{{\T}^{\Z} \}$,
$\T$ being the torus,
and $\P=\otimes_{k\in\Z}\P_k$, where $\P_{2k}=\P_0$ and $\P_{2k+1}=\P_1$ for any $k\in\Z$
are probability distributions on $\T$ and ${\cal F}$ the $\sigma$-algebra generated by the 
cylinders. We introduce
the set of random vectors on 
$(\Omega, {\cal F}, \P)$ given by 
\bea\label{beta}
& &\beta_k=(\theta_k, \alpha_k): \Omega \rightarrow \T^2,
\,\,\, k\in \Z,\nonumber\\
& &\theta_k^{\omega}=\omega_{2k}, \,\,\,\, \alpha_k^{\omega}=\omega_{2k+1}.
\eea
The random vectors $\{\beta_k\}_{k\in\Z}$ are thus i.i.d on $\T^2$.

We denote by $U_{\omega}$ the random unitary operator corresponding to the random
infinite matrix (\ref{matel}). In analogy with Jacobi matrices describing 
the discrete Schr\"odinger equation, we will also denote the vector $\ffi_k$
by the site $k$, $k\in\Z$.  

Introducing the shift operator $S$ on $\Omega$ by
\be\label{shift}
  S(\omega)_k=\omega_{k+2}, k\in\Z, 
\ee
we get an ergodic set $\{S^j\}_{j\in\Z}$ of translations.
With the unitary operator $V_j$ defined on the canonical basis of 
$l^2(\Z)$ by
\be
V_j\ffi_k=\ffi_{k-2j}, \forall k\in\Z,
\ee
we observe that for any $j\in\Z$
\be\label{ero}
U_{S^j\omega}=V_jU_{\omega}V_j^*.
\ee
Therefore, our random operator $U_{\omega}$ is a an ergodic unitary operator. 
Now, general arguments on the properties of the spectral resolution 
of ergodic operators $E_{\omega}(\Delta)$, where $\Delta$ is a Borel set of the torus
$\T$, ensure that this projector is weakly measurable, as well as 
$E^{x}_{\omega}(\Delta)=P^x_{\omega}E_{\omega}(\Delta)$, 
where $x=p.p., \;\;a.c.$ and $s.c.$,  denote the pure point, absolutely continuous
and singular continuous components, see \cite{cl}, chapter V.
The analysis performed in \cite{bhj} for the case where 
$\{(\theta_k^{\omega}, \alpha_k^{\omega})\}_{k\in\Z}$ are uniformly distributed on
the torus shows that the a.c. component of the spectrum of $U_{\omega}$ is 
almost surely empty.  

\subsection{Lyapunov Exponent}
Let us proceed by recalling some facts concerning the Lyapunov exponent.
It is shown in \cite{bb} and \cite{bhj} that generalized eigenvectors defined by
\bea\label{eveq}
&&U_{\omega}\psi=e^{i\lambda}\psi, \nonumber \\  
&&\psi=\sum_{k\in\Z}c_k\ffi_k, \,\, c_k\in \C, \,\, \lambda\in\C
\eea
in our unitary setting can be computed by means of $2\times 2$ transfer matrices
due to the structure of the matrix $U_{\omega}$. They are such that for all 
$k\in\Z$, (\cite{bhj})
\be\label{trans}
\pmatrix{c_{2k} \cr c_{2k+1}}=T(k)\pmatrix{c_{2k-2} \cr c_{2k-1}}
\ee
where the randomness lies in the phases $\eta_k(\lambda)\equiv \eta_k^{\omega}(\lambda)$ 
defined by
\be
\eta_k(\lambda)=\eta_k+\lambda,
\ee
and
\bea\label{tren}
T(k)_{11}&=& -e^{-i\eta_{2k-1}(\lambda)} \\
T(k)_{12}&=&i\frac{r}{t}\left(e^{-i\eta_{2k-1}(\lambda)}
-1\right)\nonumber\\
T(k)_{21}&=&i\frac{r}{t}\left(e^{i(\eta_{2k}(\lambda)-\eta_{2k-1}(\lambda))}
-e^{-i\eta_{2k-1}(\lambda)}\right)\nonumber\\
T(k)_{22}&=&-\frac{1}{t^2}\,e^{i\eta_{2k}(\lambda)}
+\frac{r^2}{t^2}\left(e^{i(\eta_{2k}(\lambda)-\eta_{2k-1}(\lambda))}
+1-e^{-i\eta_{2k-1}(\lambda)}\right).\nonumber
\eea
Note the properties 
\be\label{mapt}
 T(k) \equiv T(\eta_{2k}(\lambda), \eta_{2k-1}(\lambda))
\ee
whereas $ \det T(k)=e^{i(\eta_{2k}-\eta_{2k-1})}$ is independent of $\lambda$.

Therefore, knowing {\em e.g.} the coefficients $(c_0, c_1)$, we compute
for any $k\in\N$,
\bea\label{cocycle}
\pmatrix{c_{2k} \cr c_{2k+1}}&=&T(k)\cdots T(2)T(1)\pmatrix{c_{0} \cr c_{1}}
 \equiv \Phi(k)\pmatrix{c_{0} \cr c_{1}}\nonumber \\
\pmatrix{c_{-2k} \cr c_{-2k+1}}&=&T(-k+1)^{-1}\cdots 
T(-1)^{-1}T(0)^{-1}\pmatrix{c_{0} \cr c_{1}}
\equiv \Phi(-k)\pmatrix{c_{0} \cr c_{1}}.
\eea
The dynamical system at hand being ergodic and the determinant of the transfer 
matrices being of modulus one, we get the existence of a deterministic Lyapunov exponent
$\gamma(e^{i\lambda})$, for any $\lambda\in\C$, such that
\be\label{lyapu}
\lim_{k\ra\pm\infty}\frac{1}{|k|}\ln\|\Phi(k)\|=\gamma(e^{i\lambda})\;\;\; \mbox{a.s.}.
\ee
Writing $e^{i\lambda}=z\in\C\setminus\{0\}$, we also know from classical arguments, 
see {\it e.g.} \cite{cfks}, that $\gamma$ is a subharmonic function of $z$.

\setcounter{equation}{0}
\section{Density of States}

Following the standard approach in the self-adjoint case, we start by a 
definition of the density of states by averaging over the phases and invoking 
the Riesz-Markov theorem. Then we relate the density of state with alternative
definitions in terms of the density of eigenvalues of truncations of
the original operator to $l^2([M,N])$, as $N-M\ra \infty$. \\

{\bf Definition:} {\em  The density of states is the (non-random) 
measure $dk$ on $\T$ defined by 
\be\label{dos}
  \int_{\T}f(e^{i\lambda})dk(\lambda):=\E[\bra\ffi_0|f(U_{\omega})\ffi_0\ket+
\bra\ffi_1|f(U_{\omega})\ffi_1\ket
]/2,
\ee
for any continuous function $f: S^1\ra\C$.}\\
The average over the $\ffi_0$ and $\ffi_1$ matrix elements is motivated by 
the forms of the matrix (\ref{matel}) and shift (\ref{shift}). Note also that
this definition makes $dk$ a probability measure. 

Now we turn to the definition of appropriate finite size unitary matrices 
constructed from (\ref{matel}). There are several possible constructions 
suited to our purpose. Those we use below result from 
considering $U_{\omega}$ provided with  boundary conditions at certain sites
forbidding transitions through these sites.
Although such an interpretation is not needed in the sequel, let us mention
it can  be seen in \cite{bhj}. There, a more general unitary matrix 
than (\ref{matel}) is considered, whose reflection and transition coefficients 
$(r_k,t_k)$ may depend on the index $k$, whereas (\ref{matel}) is a special case 
with $r_k=r$ and $t_k=t$. Imposing $t_N=0$ there, one gets that the matrix takes 
a block structure which decouples the sites with indices smaller than $N$ from 
those with indices larger than $N$.  

Let us drop temporarily the sub- and super-scripts $\omega$ in the notation.
Fix $N\in\Z$ and consider the unitary operator $U^{2N}$ on $l^2(\Z)$
obtained from the original operator $U$ by imposing the following boundary 
conditions at the sites $2N$.
Let $U^{2N}$ be defined by (\ref{matel}) for $k\notin \{2N, 2N+1\}$ where 
\be
\eta_{2N-1}=\eta_{2N}=\eta_{2N+1}=\eta_{2N+2}=0
\ee
and, for $k\in \{2N, 2N+1\}$
\bea\label{even}
&& U^{2N}\ffi_{2N}=it\ffi_{2N-1}+r\ffi_{2N}\nonumber\\
&& U^{2N}\ffi_{2N+1}=r\ffi_{2N+1}+it\ffi_{2N+2}.  
\eea
Similarly, a boundary condition imposed at site $2N+1$ defines $U^{2N+1}$ 
by (\ref{matel}) for $k\notin \{2N, 2N+1, 2N+2, 2N+3\}$ where 
\be
\eta_{2N+1}=\eta_{2N+2}=0
\ee
and, for $k\in \{2N, 2N+1,  2N+2, 2N+3\}$
\bea\label{odd}
&& U^{2N+1}\ffi_{2N}=irte^{-i\eta_{2N}}\ffi_{2N-1}+r^2e^{-i\eta_{2N}}
\ffi_{2N}+it\ffi_{2N+1}\nonumber\\
&& U^{2N+1}\ffi_{2N+1}=-t^2e^{-i\eta_{2N}}\ffi_{2N-1}+
irte^{-i\eta_{2N}}\ffi_{2N}+r\ffi_{2N+1}\nonumber\\
&& U^{2N+1}\ffi_{2N+2}=r\ffi_{2N+2}+irte^{-i\eta_{2N+3}}\ffi_{2N+3}-
t^2e^{-i\eta_{2N+3}}\ffi_{2N+4}\nonumber\\
&& U^{2N+1}\ffi_{2N+3}=+it\ffi_{2N+2}+r^2e^{-i\eta_{2N+3}}\ffi_{2N+3}+
irte^{-i\eta_{2N+3}}\ffi_{2N+4}.  
\eea
For any $M\in\Z$, the corresponding operator $U^M$ has a the block structure 
mentionned above and it is unitary. Then, given $(M,N)\in\Z^2$ such that $M+4<N$, 
one defines a unitary matrix $U^{M,N}$ on $l^2(\Z)$ by
imposing boundary conditions at sites $M$ and $N$. By construction, $U^{M,N}$ 
contains an isolated $(N-M)\times(N-M)$ unitary block on $l^2([M+1,N])$ we
denote by $V^{M,N}$. 
Introducing the characteristic function $\chi_{M,N}$ of the set $[M+1,N]\in\Z$,
we denote by the same symbol the projector on the sites $[M+1,N]$, corresponding 
to the multiplication operator by $\chi_{M,N}$. Therefore  
\be\label{rela}
V^{M,N}=\chi_{M,N}U^{M,N}=U^{M,N}\chi_{M,N}=\chi_{M,N}U^{M,N}\chi_{M,N}.
\ee
We now consider two measures related to finite matrices as follows.\\
{\bf Definitions:} {\em  The measures $dk_{M,N}$ and $\tilde{dk}_{M,N}$ on $\T$ 
are defined by 
\bea
 && \int_{\T}f(e^{i\lambda})dk_{M,N}(\lambda):=\mbox{\em tr }
( f(V^{M,N}))/(N-M)\\
&&  \int_{\T}f(e^{i\lambda})\tilde{dk}_{M,N}(\lambda):=\mbox{\em tr }
(\chi_{M,N} f(U)\chi_{M,N})/(N-M),
\eea
for any continuous function $f: S^1\ra\C$.}\\
Note that $dk_{M,N}$ is nothing but the counting measure on $\T$ associated
with the spectrum of the finite block $V^{M,N}$, and 
$\tilde{dk}_{M,N}$ that associated with the projection of $U$ on 
$[M+1,N]$. This former operator is unitary whereas the latter is not. 

We denote the trace norm by $\|\cdot \|_1$ and 
first show a slight generalization of \cite{gt} allowing to get
\begin{lem}\label{p1} With the above notations, assume 
\be
\|(U^{M,N}-U)\chi_{M,N}\|_1=o(N-N), \mbox{ as } N-M\ra\infty,
\ee
then
\be
\lim_{N-M\ra\infty}\frac{1}{N-M}\left(\mbox{tr }( f(V^{M,N}))
-\mbox{tr }(\chi_{M,N} f(U)\chi_{M,N})\right)=0.
\ee
\end{lem}
{\bf Remark:}\\
The hypothesis is satisfied in particular if Rank$(U^{M,N}-U)<\infty$ and
uniformly bounded in $(N,M)$, as is the case with the definitions of 
$U^{M,N}$ above by means of (\ref{even}, \ref{odd})\\
{\bf Proof:}\\ 
We first note that it is enough to consider functions which are polynomials
in $z$ and $\bar{z}$, $z\in S^1$.
Any $f\in C(S^1)$ can be approximated by trigonometric polynomials
$P_R=\sum_{j=-R}^R g_je^{i j \cdot}$ in such a way that if $\eps>0$ 
is given, there exists $R(\eps)<\infty$ so that 
\be
\sup_{\theta\in\T}\left|f(\theta)-P_{R(\eps)}(\theta)\right|\leq \eps.
\ee
Hence we get  using  (\ref{rela}), 
\bea
&&\mbox{ tr }(f(V^{M,N})-\chi_{M,N}f(U)\chi_{M,N})=
\mbox{ tr }(\chi_{M,N}(f(U^{M,N})-f(U))\chi_{M,N})=\nonumber\\
&&\mbox{ tr }(\chi_{M,N}(P_{R(\eps)}(U^{M,N})-P_{R(\eps)}(U))\chi_{M,N})+\nonumber\\
&&\mbox{ tr }(\chi_{M,N}((f-P_{R(\eps)})(U^{M,N})-(f-P_{R(\eps)})(U))\chi_{M,N}),
\eea
where the trace norm of the last term is bounded by 
$2\eps (N-M)$, so that it becomes negligeable when divided by $(N-M)$.
We are thus to consider $z^s$ and $\bar{z}^s$, with $s\in\N$. 
We can write for any $s\geq 1$
\be
 U^s-(U^{N,M})^s=\sum_{j=0}^{s-1}U^j(U-U^{N,M})(U^{N,M})^{s-j-1},
\ee
so that
\be
\chi_{M,N}(U^s-(U^{N,M})^s)\chi_{M,N}=\sum_{j=0}^{s-1}\chi_{M,N}U^j
(U-U^{N,M})\chi_{M,N}(U^{N,M})^{s-j-1}.
\ee
Therefore,
\be
 \frac{\mbox{ tr }(\chi_{M,N}(U^s-(U^{N,M})^s)\chi_{M,N})}{N-M}\leq
\frac{s\|(U-U^{N,M})\chi^{M,N}\|_1}{N-M}.
\ee
The same result is true if $s<0$, with all unitaries replaced by their adjoints. 
Thus, $-R(\eps)\leq s\leq R(\eps)$ and the hypothesis on the trace norm 
of $(U-U^{N,M})\chi^{M,N}$ yield the result.\hfill \ep

Then, restoring the dependence on $\omega$ in the notation, we get by 
the same arguments as in the self adjoint case,
that the density of states is almost surely the limit in the vague sense of 
the measures $dk_{M,N}$ and $\tilde{dk}_{M,N}$ as $N-M\ra\infty$. A proof is 
provided in Appendix for completeness.

\begin{prop} \label{p2}For any continuous function $f: S^1\ra\C$,
\be\label{eqdef}
\lim_{N-M\ra\infty}\int_{\T}f(e^{i\lambda})\tilde{dk}_{M,N}^{\omega}(\lambda)=
\int_{\T}f(e^{i\lambda})dk(\lambda)\;\;\;\mbox{ a.s. },
\ee
and the support of the density of states $dk$ coincides with $\Sigma$,  the a.s. 
spectrum of $U_{\omega}$.
\end{prop}

\setcounter{equation}{0}
\section{Thouless Formula}

The link between the density of states and the Lyapunov exponent is
provided by an analysis of the spectrum of the finite unitary matrices 
$V^{M,N}$. It reads
\begin{thm}\label{ttf} [Thouless Formula ]
For any $z\in\C\setminus\{0\}$
\be\label{tf}
\gamma(z)=2\int_{\T}\ln|z-e^{i\lambda'}|dk(\lambda')+\ln(1/t^2)-\ln |z|.
\ee
\end{thm}
{\bf Remarks:}\\
0) The identity $\gamma(1/\bar{z})=\gamma(z)$ holds.
\\
i) It follows from the above formula, as in Theorem 4.6 in \cite{gt}, that the integrated 
density of states is continuous and satisfies 
\be
  |N(\lambda_1 )- N( \lambda_2)|\leq 
\frac{\ln(2/t^2)}{|\ln|e^{i\lambda_1}- e^{i\lambda_2}||}\, ,\,\,\mbox{ where }\, \, 
N(\lambda)=\int_{-\pi}^{\lambda}dk(\lambda'),
\ee
by an argument of Craig and Simon \cite{cs}.\\
ii) In case $z=e^{i\lambda}\in S^1$, the formula can be cast into the form
\be
 \gamma(e^{i\lambda})= \int_{\T}\ln(\sin^2((\lambda-\lambda')/2))dk(\lambda')+
\ln(4/t^2),
\ee 
from which we recover the estimate
$0\leq \gamma(e^{i\lambda})\leq \ln(4/t^2)$
that follows from the form of the transfer matrices (\ref{tren}).

The proof of this version of Thouless formula is given at the end of the section and
we proceed with a Corollary and an application of this formula. The Corollary 
essentially expresses the radial derivative of the Lyapunov exponent as the 
Poisson integral of the density of states measure $dk$, which allows to recover
the a.c. component of $dk$ by a limiting procedure.
\begin{cor} For any $\eps >0$ and any $\lambda'\in\T$,
\bea
&&  \lim_{\eps\ra 0^+}\gamma(e^{i\lambda'}e^{-\eps})=\gamma(e^{i\lambda'}),\\
&&  \frac{\partial}{\partial \eps} \gamma(e^{i\lambda'}e^{\pm \eps})=\mp
\int_{T}\frac{1-|e^{i\lambda'} e^{\pm \eps}|^2}{|e^{i\lambda}-
e^{i\lambda'} e^{\pm \eps}|^2}dk(\lambda)\equiv \mp 
P[dk](e^{i\lambda'} e^{\pm \eps}).
\eea
Therefore, if $n(\lambda)d\lambda/2\pi$ denotes the a.c. component 
of $dk(\lambda)$,
\be\label{limd}
\lim_{\eps\ra 0^+} \frac{\partial}{\partial \eps} 
\gamma(e^{i\lambda'}e^{-\eps})=n(\lambda')= \frac{\partial}{\partial \eps} 
\gamma(e^{i\lambda'}),
\ee 
where the limit and the derivative exist for Lebesgue almost all $\lambda'\in\T$.
\end{cor}
{\bf Remark:}\\  As in \cite{cs}, it follows also from the subharmonicity 
of $\gamma(z)$, that if $\gamma(e^{i\lambda_0})=0$, then $\gamma: S^1 \ra \R^+$ 
is continuous at  $e^{i\lambda_0}$.\\
{\bf Proof:}\\ Let us first consider the second statement with lower
indices only. We compute
\be\label{eeps}
\gamma(e^{i\lambda'}e^{-\eps})=\eps+\ln(1/t^2)+\int_{\T}
\ln(1+e^{-2\eps}-e^{-\eps}2\cos(\lambda-\lambda'))dk(\lambda),
\ee
which we can differentiate under the integral sign as long as $\eps>0$ to get
\bea
 \frac{\partial}{\partial \eps} \gamma(e^{i\lambda'}e^{-\eps})&=&1
+\int_{\T}\frac{-2e^{-2\eps}+e^{-\eps}2\cos(\lambda-\lambda')}{1+e^{-2\eps}
-e^{-\eps}2\cos(\lambda-\lambda')}dk(\lambda)\nonumber\\
&=&\int_{\T}\frac{1-e^{-2\eps}}{1+e^{-2\eps}-e^{-\eps}2\cos(\lambda-\lambda')}
dk(\lambda)=P[dk](e^{i\lambda'}e^{-\eps}).
\eea
The existence for almost all $\lambda'\in\T$ of the limit and the 
first equality in (\ref{limd}) is a direct consequence of the above equality. 
The existence and equality with the derivative at zero
for such $\lambda'$ follows from the mean value Theorem.  
To get the first statement, notice that 
$1+e^{-2\eps}-e^{-\eps}2\cos(x)>2e^{-\eps}(1-\cos(x))$ in formula (\ref{eeps}) above
yields 
\bea
&&0\leq -\ln((1+e^{-2\eps}-e^{-\eps}2\cos(\lambda-\lambda'))/4)<
-\ln(2e^{-\eps}(1-\cos(\lambda-\lambda'))/4)=\nonumber\\
&&\eps -\ln((1-\cos(\lambda-\lambda'))/2),
\eea
where the last function is in $L^1(\T, dk)$ by Thouless formula. 
Therefore, an application of the dominated convergence
Theorem shows we can take the limit
$\eps\ra 0$ inside the integral to get the result.
\hfill \ep\\

We consider now the properties of 
$U_{\omega}$ characterized by i.i.d. phases 
$\theta_k^{\omega}$ and $\alpha_k^{\omega}$ in the definition (\ref{defeta}),
assuming one set of phases is uniformly distributed on $\T$.  In that situation,
not only can we can prove the transfer matrices have a (positive) Lyapunov behaviour, but 
we can also exactly compute the Lyapunov exponent $\gamma(e^{i\lambda})$. This
shows that in this situation, the spectrum of $U_{\omega}$ is almost surely singular, 
in view of the unitary version of the Ishii-Pastur Theorem proven in \cite{bhj}.
This strengthens the corresponding results of \cite{bhj}, Theorem 4.1 and 
Propositions 5.4. There Furstenberg's Theorem is applied to prove positivity 
of the Lyapunov exponent, so that no value for $\gamma(e^{i\lambda})$ is provided.

\begin{thm}\label{co}
Let $(\theta_k^{\omega})_{k\in\Z}$ and $(\alpha_k^{\omega})_{k\in\Z}$ be i.i.d. 
on $\T$ and assume the distribution of either the $\theta_k^{\omega}$'s or the 
$\alpha_k^{\omega}$'s is uniform on $\T$. Then, for any $\lambda\in \T$,
\be
dk(\lambda)=d\lambda/2\pi,\,\,\,\mbox{ and }\,\,\,
\gamma(e^{i\lambda})=\ln(1/t^2)>0,
\ee
therefore,
\be 
 \sigma(U_{\omega})_{a.c}=\emptyset\,\,\, \mbox{ and } 
\,\,\,\sigma(U_{\omega})_{sing.}=S^1 \,\,\,\,\mbox{almost surely.}
\ee
\end{thm}
{\bf Remark:}\\
The assumption on the distribution of the phases actually implies that the $\eta_k$'s are 
i.i.d. and uniform on $T$, see Lemma \ref{uni} below. This explains why the a.s. 
spectrum coincides with $S^1$ and why the density of states is flat.\\
{\bf Proof of Theorem \ref{co}:}\\
We first use the following lemma of purely probabilistic nature proven in Appendix.
\begin{lem}\label{uni}
Under the hypotheses of Theorem \ref{co}, the $\eta_k^{\omega}$'s are i.i.d. 
and uniform on $T$.
\end{lem}
Then we show the density of states is uniform for uniformly distributed phases. 
Expanding (\ref{defeta}) of the $\eta_k(\omega)$'s 
we can write for any $n\neq 0$,
\bea \label{exploit}
&&\bra\ffi_j |U^n_{\omega}\ffi_j\ket=\sum_{\vec{k}=k_1,k_2,\cdots,k_{n-1}}(U_{\omega})_{j,k_1}
(U_{\omega})_{k_1,k_2}\cdots (U_{\omega})_{k_{n-1},j}=\nonumber\\
&&\sum_{\vec{k}}\exp\left(-i\sum_{l\in {\cal L}}p_l\eta_l(\omega)\right)(U_{0})_{j,k_1}
(U_{0})_{k_1,k_2}\cdots (U_{0})_{k_{n-1},j},
\eea
where $U_0$ corresponds to $U_{\omega}$ when all phases $\eta_k=0$ and where
${\cal L}$ is a finite set of indices depending on $j,\vec{k},n$ and $p_l$ are integers.
Observing that the variables $\eta_k(\omega)$'s all appear with the same sign
in (\ref{matel}), no compensation can take place between contributions of 
different matrix elements above and one at least among the integers $p_l$, for 
$l\in{\cal L}$ is stricly positive when $n\neq 0$. Using independence and the
characterization $\E(e^{-im\eta_k(\omega)})=\delta_{m,0}$ of the uniform distribution,
we get 
\be
\E(\bra\ffi_j |U^n_{\omega}\ffi_j\ket)=\delta_{n,0}\,\,\Longrightarrow 
\int_{\T}e^{in\lambda}dk(\lambda)=\delta_{n,0}
\ee
and the first statement follows.
The second equality is a consequence of Thouless formula together with the
identity
\be
\int_0^{2\pi}\ln|1-e^{i\lambda}|d\lambda=0.
\ee
The singular nature of the  almost sure  spectrum of $U_{\omega}$ comes from the
unitary version of Ishii-Pastur Theorem proven as Theorem 5.3 in 
\cite{bhj}, which is independent of the properties of the common distributions of the 
$\alpha_k$'s and $\theta_k$'s and only requires ergodicity. Finally, Proposition \ref{p2}
yields the result about the support of the a.s. singular spectrum.
\hfill\ep

We compute here, for the sake of completeness, the density of
states and Lyapunov exponent for the deterministic free operator $U_0$ 
corresponding to $U_{\omega}$ in case $\eta_k=0, \forall k\in\Z$. In this
case, equation (\ref{eqdef}) of Proposition \ref{p2} becomes a definition
of the free density of states $dk_0$, provided the limit exists.
That the limit exists, is the content of the next 
\begin{lem}\label{deflim} The free density of states $dk_0$ exists when defined  
for any $f\in C(S^1)$ by
\be
\int_{T}f(e^{i\lambda})dk_0(\lambda)=\lim_{N-M\ra \infty}\int_{\T}f(e^{i\lambda})
d\tilde{k}_{M,N}(\lambda).
\ee 
\end{lem}

As we know essentially everything about the purely a.c. operator $U_0$, we can also
use a direct approach to perform these computations. In particular, the integrated 
density of states of $U_0$ can be defined as the distribution function on $\T$ of 
the band functions yielding the spectrum $\Sigma_0$ of $U_0$. This direct approach 
of the density of states coincides with the above definition, 
see the proofs of 
Proposition \ref{freep} and Lemma \ref{deflim} in Appendix. 
We note here that the spectrum of $U_0$ consists in the set 
\be
\Sigma_0=\{e^{\pm i(\arccos(r^2 -t^2\cos(y)))}, y\in\T\}.
\ee
We get in particular that $\Sigma_0$ is the 
support of the density of states whereas $\Sigma_0^c$ is that of the Lyapunov exponent:
\begin{prop}\label{freep} If $N_0$, $dk_0$ and $\gamma_0$ denote the integrated density of 
states, the density 
of states and Lyapunov
exponents of $U_0$, respectively.  We have for $\lambda\in\T\simeq ]-\pi,\pi]$,
\bea
dk_0(\lambda)&=&\left\{\matrix{\frac{|\sin(\lambda)|}{2\pi\sqrt{t^4-(r^2-\cos(\lambda))^2}}d\lambda & 
\mbox{\em if }|\lambda| <\arccos(r^2-t^2)\cr 0 &  \mbox{\em  otherwise }}\right.\\
\label{nzero}
N_0(\lambda)&=&\left\{\matrix{\frac{1}{2\pi}\arccos\left(\frac{r^2-\cos(\lambda)}{t^2}\right) & 
\mbox{\em if }\lambda\in [-\arccos(r^2-t^2),0]\cr 1-\frac{1}{2\pi}
\arccos\left(\frac{r^2-\cos(\lambda)}{t^2}\right) &\mbox{\em if } \lambda\in [0,\arccos(r^2-t^2)]}\right.\\
\gamma_0(e^{i\lambda})&=&\left\{\matrix{0 & 
\mbox{\em if } |\lambda|\leq \arccos(r^2-t^2)\cr 
\cosh^{-1}\left(\frac{r^2-\cos(\lambda)}{t^2}\right)
& \mbox{\em otherwise. }}\right.
\eea
Finally, Thouless formula (\ref{tf}) holds true for these quantities with $z=e^{i\lambda}$, 
$\lambda\in\T$. 
\end{prop}
{\bf Remarks:}\\ Note that the density of $dk_0(\lambda)$  diverges as 
$1/\sqrt{|\lambda -\arccos(r^2-t^2)|}$ at the band edges
and behaves as $1/2\pi t$ as $\lambda\ra 0$. \\
The integrated density of states $N_0(\lambda)$ tends to its values $0$ and $1$ as 
$\sqrt{|\lambda -\arccos(r^2-t^2)|}$ at the band edges.\\
Also, in keeping with the fact that $U_0$ 
becomes a shift if $t=1$ and the identity as $r=1$, $N_0(\lambda)$ becomes linear in $\lambda$ as
$t\ra 1$ and a step function as $r\ra 1$.\\
The Lyapunov exponent, where non zero, is equivalently given by 
\be
\gamma_0(e^{i\lambda})=\ln \left( \frac{r^2-\cos(\lambda)}{t^2}+
\sqrt{\left(\frac{r^2-\cos(\lambda)}{t^2}\right)^2-1}\right).
\ee
It is an even $C^{\infty}$ function of $\lambda$ on $\{ |\lambda|> \arccos(r^2-t^2) \}$, 
strictly increasing 
on $[\arccos(r^2-t^2),\pi]$. And $d \gamma_0(e^{i\lambda})/d\lambda$ behaves as 
$1/\sqrt{\lambda- \arccos(r^2-t^2)}$ as $\lambda\ra  \arccos(r^2-t^2)^+$.\\
Given Lemma \ref{deflim} above, it is clear that Thouless formula holds
for the above quantities. A direct proof of this fact is nevertheless given in Appendix.

\subsection{Proof of Thouless Formula}
We now turn to the proof of Theorem \ref{ttf}.
Writing down explicitely the effect of the boundary conditions 
at $N>M$ on the coefficients of the eigenvector (\ref{eveq}) we obtain 
the following relations, which depend on the parity of $N$ and $M$.
Let $\psi^{M,N}=\chi_{M,N}\psi$ and consider 
\be\label{redeq}
V^{M,N}\psi^{M,N}=e^{i\lambda}\psi^{M,N} \,\,\,\, \mbox{ in } l^{2}[M+1,N].
\ee
We get by inspection,
\begin{lem}  Assume (\ref{redeq}) is satisfied. Then, if $M$ is even
\be
 \pmatrix{c_{M+2}\cr c_{M+3}}=c_{M+1}b_1(e^{i\lambda})\equiv
c_{M+1} \frac{1}{t^2}\pmatrix{-it(r-e^{-i\lambda})
\cr (r-e^{i\lambda})+r(r-e^{-i\lambda})}.
\ee
If $M$ is odd, 
\be
 \pmatrix{c_{M+1}\cr c_{M+2}}=c_{M+1}b_2(e^{i\lambda})\equiv
c_{M+1}\frac{1}{it}\pmatrix{it \cr e^{i\lambda}-r}. 
\ee
Similarly, if $N$ is even,
\be
 \pmatrix{c_{N-2}\cr c_{N-1}}=c_{N}b_3(e^{i\lambda})\equiv
c_{N} \frac{1}{t^2}\pmatrix{
 (r-e^{i\lambda})+r(r-e^{-i\lambda})\cr -it(r-e^{-i\lambda})}.
\ee
If $N$ is odd,
\be
 \pmatrix{c_{N-1}\cr c_{N}}=c_{N-1}b_4(e^{i\lambda})\equiv
c_{N-1}\frac{1}{it}\pmatrix{e^{i\lambda}-r \cr it}. 
\ee
\end{lem}
These relations together with the formulas (\ref{cocycle}) allow 
to describe the spectrum of $V^{M,N}$ in a convenient manner.
\begin{cor}\label{corol}
Let $M<N$ be fixed and consider non zero vectors $a_1, a_2\in\C^2$ 
such that $a_j(e^{i\lambda})\in (b_{j+2}(e^{i\lambda})\C)^{\perp}$,
$j=1,2$. 
Then, $ e^{i\lambda}\in\sigma(V^{M,N}) $ iff 
\bea
&&\bra a_1(e^{i\lambda})|T(N/2-1)\cdots 
T(M/2+2)b_1(e^{i\lambda})\ket=0, \hspace{2.25cm} M,N \mbox{ even }\nonumber\\
&&\bra a_2(e^{i\lambda})|T((N+1)/2-1)\cdots 
T(M/2+2)b_1(e^{i\lambda})\ket=0,  \hspace{.4cm} M \mbox{ even }, 
N \mbox{ odd }\nonumber\\
&&\bra a_1(e^{i\lambda})|T(N/2-1)\cdots 
T((M+1)/2+1)b_2(e^{i\lambda})\ket=0, \hspace{.4cm}  M \mbox{ odd }, 
N \mbox{ even }\nonumber\\
&&\bra a_2(e^{i\lambda})|T((N+1)/2-1)\cdots 
T((M+1)/2+1)b_2(e^{i\lambda})\ket=0, \hspace{.46cm}  M,N \mbox{ odd }
\eea
\end{cor}
{\bf Remark:}\\ In particular, a possible choice for the $a_j$' is
\be\label{adb}
 a_1(e^{i\lambda})=b_1(e^{-i\lambda}), \,\,\, a_2(e^{i\lambda})=b_2(e^{-i\lambda}).
\ee
Each of the above quantities denotes a matrix element of 
a product of transfer matrices of the type (\ref{cocycle}), which 
depend on $e^{i\lambda}$, and will be linked in the limit 
$N-M\ra \infty$ to the Lyapunov exponent.

Let $e^{i\lambda}=z\in\C\setminus\{0\}$ and $n_0, m_0\in \Z$. Defining 
\be
\Phi^{m_0,n_0}(z)=T(n_0-1)\cdots T(m_0+2),
\ee
one sees that the matrix elements $\bra a_j(z)|\Phi^{m_0,n_0}(z)b_k(z)\ket$
correspond to those in the above corollary for values $N=2n_0, N=2n_0-1, M=2m_0, M=2m_0+1$,
depending on the choice of indices $j,k$.
\begin{lem}  For any  $z\in\C\setminus S^1$ and any indices $j,k=1,2$ 
\bea\label{rhs}
&&\lim_{n_0-m_0\ra \infty}\frac{1}{2(n_0-m_0)}\ln|\bra a_j(z)|
\Phi^{m_0, n_0}(z)b_k(z)\ket|=\nonumber\\
&&\hspace{4cm}\int_{\T}\ln|z-e^{i\lambda'}|dk(\lambda')+\ln(1/t)-\ln(|z|^{1/2}),
\eea
\end{lem}
{\bf Proof:}
We note that for any $k\in\Z$, there exist $2\times 2$ matrices $A(k), B(k), C(k)$ 
such that (with $z=e^{i\lambda}$)
\be
T(k)=zA(k)+B(k)+C(k)/z,\,\, \mbox{ where } 
A(k)=\pmatrix{0&0\cr 0& -\frac{-e^{i\eta_{2k}}}{t^2}}
\ee
Also, for any $j=1,2$, there exist vectors $b_j^{(k)}$, $a_j^{(k)}$, $k=-1,0,1$ such that
\bea
 a_k(z)&=&za_k^{(1)}+a_k^{(0)}+a_k^{(-1)}/z, \nonumber\\
 b_k(z)&=&zb_k^{(1)}+b_k^{(0)}+b_k^{(-1)}/z, 
\eea
where $b_2^{(-1)}=a_2^{(1)}=0$ are the only non zero vectors with the choice (\ref{adb}).
Thus, taking into account the above property ,
\be
P_{j,k}(z)=z^{n_0-m_0+(1-k)}\bra a_j(z)|\Phi^{m_0, n_0}(z)b_k(z)\ket
\ee
is a polynomial in $z$ of degree $2(n_0-m_0)+2-(k+j)$. 
Let $p_{j,k}$ be the coefficient of the highest
power of $z$  of $P_{j,k}$. Then, because of corollary \ref{corol}, we can write
\be
  P_{j,k}(z)=p_{j,k}\prod_{l=0}^{\mbox {\scriptsize deg }P_{j,k}}(z-e^{i\lambda_l}),
\ee
where $\{e^{i\lambda_l}\}$ is the set of eigenvalues of $V^{M,N}$ and we compute
\bea
&&|p_{j,k}|=|\bra a_j^{(2-j)}|\prod_{l=m_0+2}^{n_0-1}A(l)b_k^{(1)}\ket|=\nonumber\\
&&\frac{K_0}{t^{2(n_0-m_0)}}
\left|{\bigg \langle }\pmatrix{-it\cr r}{\bigg |}\pmatrix{0&0\cr 0&1}^{(n_0-m_0)-2}
\pmatrix{0\cr 1}{\bigg \rangle}\right|=\frac{K_1}{t^{2(n_0-m_0)}}
\eea
where $K_0, K_1$ are some constants that depends on $j,k$ and $t$.
Therefore, for any $z\in \C\setminus S^1$,
\be
\lim_{n_0-m_0\ra \infty}\frac{\ln|P_{j,k}(z)|}{(n_0-m_0)}=\ln(1/t^2)+
\lim_{n_0-m_0\ra \infty}\sum_{l=0}^{\mbox {\scriptsize deg }P_{j,k}}
\frac{\ln|z-e^{i\lambda_l}|}{(n_0-m_0)}
\ee
Introducing the continuous function $f_z: S^1\ra\R$ given by $f_z(x)=\ln|z-x|$,
the last term can be written
\be
  \lim_{n_0-m_0\ra \infty }
\sum_{l=0}^{\mbox {\scriptsize deg }P_{j,k}}\frac{f_z(e^{i\lambda_j})}{n_0-m_0}=
2\lim_{M-N\ra\infty}\frac{\mbox{ tr }(f_z(V^{M,N}))}{N-M}=
2\int_{\T}f_z(e^{i\lambda'})dk(\lambda')
\ee
by application of Lemma \ref{p1} and Proposition \ref{p2}. 
This ends the proof of the lemma.  
\hfill \ep

Then we make use the following easy lemma
\begin{lem}
If $\Phi: \C^2\ra\C^2$ is linear and $a_j, b_j\in\C^2$, $j=1,2$ are such 
that \,$\mbox{\em span }(a_1,a_2)=\mbox{\em span }(b_1,b_2)=\C^2$, then 
$\|\Phi\|:=\max_{j,k}|\bra a_j |\Phi b_k\ket|$ is a norm for $\Phi$,
\end{lem}
noting that its hypothesis is satisfied by $a_k(z), b_j(z)$, for all  
$z\neq -1$, 
and of the fact that the Lyapunov exponent is defined independently of the 
norm used in (\ref{lyapu}) to deduce that (\ref{rhs}) actually equals half 
the Lyapunov exponent. 
Finally, the fact that both the Lyapunov exponent
and the right hand side of (\ref{rhs}) are subharmonic and coincide on
$\C\setminus S^1$ implies the relation (\ref{tf}) on $\C$ as well, 
by classical arguments, see \cite{cs}. This ends the proof of Thouless formula.\hfill\ep

\setcounter{equation}{0}
\section{Properties of the Density of States}

We mentionned several times the analogy between our unitary operator $U_{\omega}$ and Jacobi matrices
corresponding to the self-adjoint case. In this section we slightly drift away from the physical 
motivations underlying the study of (\ref{matel}) and consider more closely the links between
these cases. The analogy is made clearer by the following Lemma which will be useful later.
\begin{lem}\label{anderson} Denoting unitary equivalence by $\simeq$, we have 
\be
 U_{\omega}\simeq D_{\omega}S_0, \,\,\,\mbox{ with } 
D_{\omega}=\mbox{\em diag }\{e^{-i\eta_k^{\omega}}\}
\ee
and 
\be
S_0=\pmatrix{\ddots & rt & -t^2& & & \cr
              & r^2& -rt  & & & \cr
              & rt & r^2 & rt & -t^2& \cr
              & -t^2 &-tr & r^2& -rt& \cr 
              & & & rt &r^2 & \cr
              & & & -t^2& -tr&\ddots }\simeq U_0,
\ee 
where the translation along the diagonal is fixed by 
$\bra \ffi_{2k-2}|S_0\ffi_{2k}\ket =-t^2$, $k\in\Z$.
\end{lem}
{\bf Remarks:}\\
In some sense, the Lemma says that, 
up to unitary equivalence, $U_{\omega}$ is a unitary analog of the one dimensional discrete random 
Schr\"odinger operator where the a.c. unitary $S_0$ plays the role of the discrete laplacian, 
the pure point diagonal operator $D_{\omega}$ plays the role of the potential on the sites, and the operator 
sum is replaced by a product.\\
We also recall that tridiagonal unitary matrices are spectrally uninteresting as they either correspond
to a shift of to infinite direct sums of blocks of size one or two, see Lemma 3.1 in \cite{bhj}.\\ 
The Lemma also shows that our operator $U_{\omega}$ is essentially a product
of an absolutely continuous unitary and a pure point unitary, whereas it was constructed in 
Section 2 of \cite{bhj} as a product of two pure point unitaries. \\
{\bf Proof:}\\
Let us define a collection of rank two operators by 
\be
P_j=|\ffi_j\ket\bra \ffi_j|+|\ffi_{j+1}\ket\bra \ffi_{j+1}| , \,\,\, j\in \Z,
\ee
and the unitary $V$ by the direct sum
\be
V={\sum_{j\in\Z}}^{\oplus}P_{2j-1}\pmatrix{ir&t\cr -it&r}P_{2j-1}.
\ee
It is just a matter of computation to check that we can write
\be
U_{\omega}=(U_{\omega}U_0^{-1})U_0\equiv V^{-1}D_{\omega}VU_0=V^{-1}D_{\omega}(VU_0V^{-1})V
\equiv V^{-1}(D_{\omega}S_0)V,
\ee
with the required properties for $S_0$ and $D_{\omega}$.\hfill \ep\\

Now, forgetting that the phases $\eta_k^{\omega}$ are in general correlated random variables,
see (\ref{defeta}), if we consider them as i.i.d., but not necessarily uniformly distributed on $\T$,
we get some unitary Anderson-like model. This is where we depart from the physical motivation, as
it is recalled in Lemma 4.2 in \cite{bhj} that independence of the $\eta_k$'s is associated with 
a uniform distribution. 

\subsection{Support of the Density of States}

Nevertheless, assuming the random phases $\{\eta^{\omega}_k\}_{k\in\Z}$ are i.i.d. according to the
measure $d\mu$ on $\T$, we can characterize the almost sure spectrum 
of $U_{\omega}$ in term of the support of $\mu$ and of the spectrum $\Sigma_0$ of $U_0$.
\begin{thm} Under the above hypotheses, 
the almost sure spectrum of $U_{\omega}$ consists in the set
\be
\Sigma:=\exp(i\,\mbox{\em supp} \mu)\Sigma_0=
\{e^{i\alpha}\Sigma_0\,\,| \,\,\alpha\in\mbox{\em supp} \mu \}.
\ee
\end{thm}
{\bf Remark:}\\ In the case where the $\eta_k(\omega)$ are i.i.d. and uniform on $\T$,
we recover the fact that the almost sure spectrum of $U_{\omega}$ is $S^1$.\\
{\bf Proof:}\\ To show that $\Sigma$ belongs to the almost sure spectrum, we simply 
construct Weyl sequences corresponding to the
corresponding quasi-energies, with probability one.
We know from Section 6 of \cite{bhj} that for any $e^{i\lambda} \in\Sigma_0$, there exists 
a generalized eigenvector $\psi_{\lambda}$ such 
\be\label{propgen}
\psi_{\lambda}=\sum_{j\in\Z}c_j(\lambda)\ffi_j, \,\, 
U_0\psi_{\lambda}=e^{i\lambda}\psi_{\lambda},\mbox{ and } \,\, 
 0<K<|c_j(\lambda)|<1/K, \,\,\forall j\in\Z,
\ee
for some $K>0$. The last property can be checked also by means of the 
transfer matrices (\ref{tren})

Let $\alpha\in \mbox{supp} \mu$. Then, for all  $\eps >0$, there exists a set 
$I_{\eps}\ni \alpha$ such that $|I_{\eps}|\leq \eps$,  and $\mu( I_{\eps})>0$.
With the notation $\omega(k)=\eta_k(\omega)$, $k\in\Z$, we define for all $n\in\N$ and 
$k\in\Z$, 
\be
A_n(k)=\{\omega(kn)\in I_{\eps},\omega(kn+1)\in I_{\eps}, \cdots ,
\omega(kn+ n-1)\in I_{\eps}  \}.
\ee
Due to the assumed independence, we have for any $k$, 
$ \P(A_n(k))=\mu( I_{\eps})^n >0$
so that for any $n>0$, by Borel-Cantelli, 
$\P(\cup_{k\in \Z}A_n(k))=1$.

Let $\Delta_n(k)=\{kn, kn+1, \cdots, kn+n-1\}$ denote the set of indices appearing in 
$A_n(k)$ and consider now
\be
\psi_{n,k}(\lambda)=\sum_{j\in\Delta_n(k)}c_j(\lambda)\ffi_j=
\chi(\Delta_n(k))\psi(\lambda),
\ee where $\chi(\Delta_n(k))$ is the projector on the span of $\{\ffi_j\}_{j\in \Delta_n(k)}$
Because of (\ref{propgen}), 
\be
U_0\psi_{n,k}(\lambda)=e^{i\lambda}\psi_{n,k}(\lambda) + R^-_{kn}(\lambda)+R^+_{k(n+1)}, 
\ee
where the vectors $R^{\pm}_{j}$ have at most four components close to the index $j$ and
\be
\|R^{\pm}_{j}\|\leq R, \mbox{ where } R \mbox{ is uniform in }  j.
\ee
Also, by construction of $A_n(k)$, $U_0$ and $U_{\omega}$, we have
\bea
\| U_{\omega}\psi_{n,k}(\lambda)-e^{i\alpha}U_0\psi_{n,k}(\lambda)\|
&\leq &\| (U_{\omega}-e^{i\alpha}U_0)
\chi(\Delta_n(k))| \|\psi_{n,k}(\lambda)\|\nonumber\\
& = &O(\eps) \|\psi_{n,k}(\lambda)\|,
\eea
where the estimate  $O(\eps)$ is uniform in $n$ and $ k$. Therefore, for all $\eps>0$
and all $n>0$, there exists, with probability one, a $k$ such that $A_n(k)$ and the 
corresponding $\psi_{n,k}(\lambda)$ have the above properties so that 
\bea
&&\|U_{\omega}\psi_{n,k}(\lambda)-e^{i(\alpha+\lambda)}\psi_{n,k}(\lambda)\|/
\|\psi_{n,k}(\lambda)\|=\nonumber\\
&&(\|(U_{\omega}-e^{i\alpha}U_0)\psi_{n,k}(\lambda)+
e^{i\alpha}(U_0-e^{i\lambda})\psi_{n,k}(\lambda)\|)/\|\psi_{n,k}(\lambda)\|\leq \nonumber\\
&& O(\eps)+2R/\|\psi_{n,k}(\lambda)\|=O(\eps + 1/n).
\eea
It remains to chose $n= [1/\eps]$ to conclude that $e^{i(\alpha+\lambda)}\in
\sigma(U_{\omega})$ almost surely.

Let us now show that $S^1\setminus \Sigma$ belongs to the resolvent set of 
$U_{\omega}$. In order to do so we use Lemma \ref{anderson}
Therefore, we can consider as well the spectrum of the product $D_{\omega}S_0$
to which the perturbation theory recalled in Chap.1, \S 11 of \cite{yaf} for example,
applies. In particular, dropping the $\omega$ in the notation as randomness plays 
no role here, if we know that for all $j\in\Z$, $\eta_j\in[\alpha,\beta]\subset \T$, then 
$\sigma(D)\subseteq (\delta_1,\delta_2)$ where  $(\delta_1,\delta_2)$ denotes the corresponding 
arc on the unit circle swept in the positive direction from $\delta_1\in S^1$ to 
$\delta_2\in S^1$. We denote by $|(\delta_1,\delta_2)|$ the length on the torus of this 
arc. Since $\sigma(S_0)=\Sigma_0$ corresponds to the symmetric arc $(e^{-i\arccos(r^2-t^2)}, 
e^{i\arccos(r^2-t^2)})$, perturbation theory tells us that after (multiplicative) 
perturbation by $S_0$, the spectrum of $U\simeq DS_0$ is a subset of an arc of wider aperture than
$(\delta_1,\delta_2)$. Quantitatively, Theorem 8, p.65 in \cite{yaf} tells us that the arc 
$(e^{i\arccos(r^2-t^2)}\delta_2,e^{-i\arccos(r^2-t^2)}\delta_1)$ belongs to the 
resolvent set of $U$, provided $|(\delta_1,\delta_2)|<|(e^{i\arccos(r^2-t^2)}, 
e^{-i\arccos(r^2-t^2)})|$. This condition simply insures that the subset of the resolvent 
set we are talking about is not reduced to the empty set. This is enough to get the result
in case the support of $\mu$ is such that $\Sigma$ is connected. In case this set is not 
connected, as $|\Sigma_0|>0$, it consists of a finite set of connected components, each 
of which can be associated with the convex hull of sufficiently far apart subsets of
the support of $\mu$. Denoting these subsets by $m_j$, $j=1,\cdots,N$ and the associated 
arcs on $S^1$ by $(M_1(j),M_2(j))$, we have that the spectrum of $D$ is the disjoint union
of subsets $\sigma_j$ satisfying $\sigma_j\subseteq (M_1(j),M_2(j))$. The same argument as above
says that the spectrum of $DS_0$ is confined to the finite union of arcs 
$((e^{i\arccos(r^2-t^2)}M_1(j),(e^{-i\arccos(r^2-t^2)}M_2(j))$, which ends the 
proof of the Theorem.
\hfill \ep

\subsection{Analyticity of the density of states}
At the price of some combinatorics, we can further exploit the relation 
(\ref{exploit}) in order to 
obtain a condition on the common distribution of the $\eta_k$'s ensuring the analyticity 
of the density of states. Recall that a function $f$ on $\T$ is analytic, if and only if 
its Fourier coefficients $\hat{f}$ satisfy an estimate of the form
\be\label{four}
 |\hat{f}(n)|\leq A e^{-B|n|}, \,\,\, \forall n\in\Z,
\ee
for some positive constants $A,B$. We have
\begin{thm}\label{52} Assume the $\eta_k$'s are distributed according to a law that 
has an 
analytic density $f$ characterized by the estimate (\ref{four}) with $A,B>0$. 
Then, if 
\be
B>\ln(1+2rt)+\ln A,
\ee
the density of states $dk$ admits an analytic density, so that the integrated 
density of states $N$ is analytic as well.
\end{thm}
{\bf Remarks:}\\ As $\hat{f}(0)=\int_{\T}f(\eta)d\eta =1$, $A\geq 1$.\\
When the Theorem applies, it prevents the Lyapunov exponent from being zero
on a set of positive measure.\\
This result has to be compared with the Proposition VI. 3.1. of \cite{cl} stating
a similar result for the $d$-dimensional Anderson model.\\

As an immediate consequence, using $r^2+t^2=1$, we get the following
\begin{cor}
If the $\eta_k$'s have an analytic density $f$, characterized by
(\ref{four}) with $B>\ln A$, then there exist $r^+(f)$ and $r^-(f)$ in
$]0,1[$ such that the density of states is analytic provided the reflexion coefficient
$r$ satisfies $1>r>r^+(f)$ or $0<r<r^-(f)$. If $B>\ln(2A)$, The density of state 
is analytic $\forall r\in[0,1]$.  
\end{cor}
{\bf Remark:}\\ It is easy to check that in both the extreme cases $r=1$ and $r=0$,
the density of states is analytic. Indeed, if $r=1$, $dk(\lambda)=f(\lambda)d\lambda$, where
$f$ is the density of the $\eta_k$'s, whereas if $r=0$, $dk(\lambda)=d\lambda /(2\pi)$.\\
{\bf Proof of Theorem \ref{52}:}\\
By hypothesis, for any $n\in\Z$, 
\be
  |\Phi_{\eta}(n)|=\left|\int_{\T}e^{i\eta n}f(\eta)d\eta\right|\leq  A e^{-B|n|}.
\ee
Then, in (\ref{exploit}) above, $\sum_{l\in {\cal L}}p_l=n$, so that using independence
\be\label{rhss}
|\E\bra\ffi_j |U^n_{\omega}\ffi_j\ket |\leq A^ne^{-Bn}
\left.\sum_{k_1,k_2,\cdots,k_{n-1}}|(U_{0})_{j,k_1}|
|(U_{0})_{k_1,k_2}|\cdots |(U_{0})_{k_{n-1},j}|\right.
\ee
Here the sum carries over a set of indices that form paths of length $n+1$ 
from index $j$ to index $j$. The allowed paths are those giving rise to non zero 
matrix elements $(U_0)_{l,m}$ in the sum above. In order
to compute this last sum, we proceed as follows. Let us 
introduce more general $j$-dependent subsets ${\cal C}_{n-1}(j)$ of indices of $\Z^{n-1}$
that appear in the computation of the matrix element $\bra\ffi_0|U_{\omega}^n\ffi_j\ket$.
This set consists of paths of the form
$\{k_0=0, k_1, k_2, \cdots, k_{n-1}, k_n=j\}$ of length $n+1$ in $\Z$ from $0$ to $j$ 
with the condition that 
\bea
&& k_{m+1}-k_m\in\{0,+1,-1,+2\} \quad\mbox{ if }  \quad k_{m} \mbox{ is odd}\nonumber\\
&& k_{m+1}-k_m\in\{0,+1,-1,-2\} \quad\mbox{ if } \quad k_{m} \mbox{ is even},
\eea
for all $m=0,1,\cdot, n-1$. Let us define 
\be
S_{n-1}(j):=\sum_{{\cal C}_{n-1}(0)}|(U_{0})_{0,k_1}|
|(U_{0})_{k_1,k_2}|\cdots |(U_{0})_{k_{n-1},j}|,
\ee
where the matrix elements $|(U_{0})_{l,m}|$ are given by $r^2, rt$ and $t^2$ respectively, 
when $|l-m|$ equals $0$, $1$ and $2$ respectively. 
This quantity actually gives a crude upper bound on the probability to go from
site $0$ to $j$ in $n$ time steps, under the free evolution. It is crude in the sense
that it does not take the phases into account during that free evolution.
 
We are actually interested in the computation of $S_{n-1}(0)$ and of the similar quantity 
appearing in the computation of  
$\bra\ffi_1|U_{\omega}^n\ffi_1\ket$, which correspond the the sum in the right hand side
of (\ref{rhss}), in the asymptotic regime $n\ra\infty$. The case of the matrix element
$\bra\ffi_1|U_{\omega}^n\ffi_1\ket$ being similar, we only consider $S_{n-1}(0)$.

The plan is to use a transfer matrix formalism to evaluate the generating 
function associated with $S_{n-1}(j)$ and then to compute the asymptotics of 
$S_{n-1}(0)$. 
In view of (\ref{rhss}), the following proposition implies the Theorem.
\begin{prop} \label{comb} For some constant $c>0$,
\begin{eqnarray}  
 S_{n-1}(0)= \frac{c(r+t)^{2n}}{\sqrt{n}}(1+o(1)) \,\,\,\mbox{ as }\, n\ra\infty .
\end{eqnarray}
\end{prop}
{\bf Proof of Proposition \ref{comb}:}\\
Let 
\be
P_n(x)=\sum_{-2n\leq j\leq 2n}{S}_{n-1}(j)x^j
\ee 
be this generating function which we split into two parts 
$P_n(x)=P_n^+(x)+P_n^-(x)$ where
\be
P_n^{\pm}(x)=\sum_{\scriptsize \matrix{-2n\leq j \leq 2n \cr
j \matrix{ \mbox{ even}\cr\mbox{ odd}} }}{S}_{n-1}(j)x^j.
\ee
Clearly we have for $n=0, 1$,
\be
P_0^+(x)=r^2, P_0^-(x)=0, P_1^+(x)=r^2+t^2x^{-2}, P_1^-(x)=rt(x+x^{-1}). 
\ee
It is readily shown by induction that a transfer matrix allows
to compute $P_n(x)$  for any $n$: 
\begin{lem}\label{tramat}
For any $n\geq0$,
$$
\pmatrix{P_{n+1}^+(x)\cr P_{n+1}^-(x)}=\pmatrix{r^2+t^2x^{-2} & rt(x+x^{-1}) \cr
rt(x+x^{-1}) & r^2+t^2x^{2}}\pmatrix{P_n^+(x)\cr P_n^-(x)},
$$
with $P_0^+(x)=r^2, P_0^-(x)=0$.
\end{lem}
Denoting by $T(x)$ the transfer matrix defined in this Lemma, and introducing the
parameter 
\be
\tau=t/r \in ]0,\infty[,
\ee
we rewrite it as
\be
T(x)=r^2\pmatrix{1+\tau^2x^{-2} & \tau(x+x^{-1}) \cr
\tau(x+x^{-1}) & 1+\tau^2x^{2}}.
\ee
We will consider first the case $t\neq r \Longleftrightarrow \tau\neq 1$. The
case $\tau=1$, for which more can be said about $S_{n-1}(j)$, see Proposition 
\ref{better}, is dealt with below.
\subsubsection{Case $\tau\neq 1$}
The eigenvalues of $T(x)$ are given by $r^2$ times $\lambda_{\pm}(x)$, where
\be
  \lambda_{\pm}(x)=\left\{1+\tau(x^2+x^{-2})/2\pm
\sqrt{(1+\tau(x^2+x^{-2})/2)^2-(1-\tau^2)^2}\right\},
\ee
so that
\be
T^n(x)=r^{2n}A(x)\pmatrix{\lambda_{+}^n(x)& 0 \cr 0 & \lambda_{-}(x)^{n}}A(x)^{-1}
\ee
with 
\be
A(x)=\pmatrix{\lambda_+(x)-(1+\tau^2 x^{2}) & \lambda_-(x)-(1+\tau^2 x^{2})\cr
\tau(x+x^{-1}) & \tau(x+x^{-1})}.
\ee
For the moment, $x$ is just book keeping parameter, so that we ignore the potential 
problems of the definition of $A(x)$ in case the eigenvalues are degenerate
and we further compute
\bea
&&\pmatrix{P_n^+(x)\cr P_n^-(x)}=T^n(x)\pmatrix{r^2\cr 0}=\\
&&\frac{r^{2n}\tau(x+x^{-1})}
{2\sqrt{(1+\tau(x^2+x^{-2})/2)^2-(1-\tau^2)^2}}\times\nonumber\\
&&\quad\quad \quad\quad\quad\quad\pmatrix{\lambda_+(x)^{n+1}-\lambda_-(x)^{n+1}-(\lambda_+(x)^n-\lambda_-(x)^n)
(1+\tau^2x^2)\cr \tau(x+x^{-1})(\lambda_+(x)^n-\lambda_-(x)^n)}.\nonumber
\eea
We note at this point that one checks, using the binomial Theorem, that
despite the presence of square roots in the expressions for $P_n^{\pm}(x)$,
these quantities actually are given by finite Laurent expansions in $x$, as they 
should. Focusing on $P_n^+(x)$ we can rewrite with the shorthand $\sqrt{\cdot}$ 
for the square root of the denominator above
\bea
&&  P_n^+(x)=\\
&& \quad \frac{r^{2n}\tau(x+x^{-1})}
{2\sqrt{\cdot}}\left((\lambda_+(x)^n-\lambda_-(x)^n)
\frac{\tau^2}{2}(x^{-2}+x^2)+\frac{\sqrt{\cdot}}{2}
(\lambda_+(x)^n+\lambda_-(x)^n)\right).\nonumber
\eea  
The quantity of interest to us is $S_{n-1}(0)$, the coefficient of
$x^0$ in the expansion of  $ P_n^+(x)$. Substituting $e^{i\theta}$ for
$x$ in $P_n^+$, we get a trigonometric polynomial whose zero'th Fourier
coefficient is obtained by integration
\be\label{fourier}
  S_{n-1}(0)=\int_{\T}P_n^+(e^{i\theta})d\theta/(2\pi).
\ee
It remains to perform the asymptotic analysis of the above integral as
$n\ra\infty$. It is a matter of routine
to verify the following propereties: The eigenvalues, as functions of 
$\theta\in \T\simeq ]-\pi, \pi]$, are
continuous. If $\tau<1$, they are real valued, with discontinuity 
of the derivative at $\theta=\pm\pi/2$, where they cross and are given by
$1-\tau^2$. At all other values of $\theta$, they are $C^{\infty}$ and 
they satisfy 
\be\label{satis}
 \lambda_+(e^{i\theta})>\lambda_-(e^{i\theta}), \mbox{ with }
\lambda_+(e^{i\theta})>1-\tau^2.
\ee
If $\tau>1$, the eigenvalues become complex conjugate. Let 
$\theta_c=\arccos(\frac{\tau^2-2}{\tau^2})/2$ be the critical value where
the square root becomes zero. If 
$\theta \in [\theta_c,\pi-\theta_c]\cup [-\pi+\theta_c, -\theta_c]$,  the
eigenvalues are complex conjugate, of modulus $|1-\tau^2|$. Otherwise they
are real valued, and satisfy (\ref{satis}) as well. Therefore, the asymptotics 
as $n\ra\infty$ of (\ref{fourier}) is determined by $\lambda_+$ only. 
Moreover, in both cases, $\ln(\lambda_+(e^{i\theta}))$ admits 
non degenerate maxima at $\theta=0$ and $\pi$, where $\lambda_+$ reaches its
 maximum value $(1+\tau^2)$.
Therefore, Laplace's method yields the asymptotics of the Proposition. \hfill\ep

\subsubsection{case $\tau=1$}

The course of the proof being the same, it is presented in Appendix. 
However, instead of computing
$S_{n-1}(0)$ as $n\ra\infty$, we can get exact forms for all $S_{n-1}(j)$'s. The
Proposition we actually show is 
\begin{prop}\label{better}
\begin{eqnarray}
 { S}_{n-1}(j)&=&\frac{1}{2^n}\pmatrix{2n-1 \cr j/2+n}, 
\quad \quad -2n\leq j\leq 2(n-1), \quad \quad j \mbox{ \em even}\nonumber\\
{S}_{n-1}(j)&=&\frac{1}{2^n}\pmatrix{2n-1 \cr (j-1)/2+n}, 
\quad \quad -2n+1\leq j\leq 2n-1, \quad \quad j \mbox{ \em  odd}
\end{eqnarray}
{\bf Remark:}\\ Of course, Stirling's formula for  $n$ large yields 
proposition \ref{comb} with $r=t=1/\sqrt{2}$:
\be
  S_{n-1}(0)=\frac{1}{2^{n}}\pmatrix{2n-1 \cr n}\simeq \frac{2^{n}}{\sqrt{\pi n}}.
\ee
\end{prop}

\setcounter{equation}{0}
\section{Appendix}

 {\bf Proof of Proposition \ref{p2}:}\\
 We have by definition,
 \be
 \int_{\T}f(e^{i\lambda})\tilde{dk}_{M,N}^{\omega}(\lambda)=\frac{1}{N-M}\sum_{j=M+1}^N\bra 
 \ffi_j |f(U_{\omega})\ffi_j\ket,
 \ee
 where, depending on the parity of $M$ and $N$ and due to the fact that $f$ is uniformly bounded, 
 the right hand side can be rewritten as
 \bea
 && \frac{1}{N-M}\left(\sum_{k=(M+1)/2}^{N/2}
 \bra \ffi_{2k} |f(U_{\omega})\ffi_{2k}\ket+\bra \ffi_{2k+1} |f(U_{\omega})\ffi_{2k+1}\ket\right)
 +O_f(\frac{1}{N-M})=\nonumber\\
 &&\frac{1}{N-M}\left(\sum_{k=(M+1)/2}^{N/2}
 \bra \ffi_{0} |f(U_{S^k(\omega)})\ffi_{0}\ket+\bra \ffi_{1} |f(U_{S^k(\omega)})\ffi_{1}\ket\right)
 +O_f(\frac{1}{N-M}).
 \eea
 Now, by  Birkhoff ergodic theorem, there  exists $\Omega_f$ of measure one such that for all
 $\omega\in\Omega_f$, 
 \be
 \lim_{N-M\ra \infty}\frac{1}{N-M}\sum_{k=(M+1)/2}^{N/2}\bra \ffi_{j} |f(U_{S^k(\omega)})\ffi_{j}\ket=
 \frac{1}{2}\E(\bra \ffi_{j} |f(U_{\omega})\ffi_{j}\ket) , \forall j\in\Z,
 \ee
 therefore,
 \be
 \frac{1}{N-M}\mbox{ tr } (\chi^{M,N}f(U_{\omega}))\ra \frac{1}{2}
 \left(\E(\bra \ffi_{0} |f(U_{\omega})\ffi_{0}\ket+\bra \ffi_{0} |f(U_{\omega})\ffi_{0}\ket)\right).
 \ee
 Then, $C(S^1)$ being separable, we have the existence of a countable set of $\{f_j\}_{j\in\N}$, dense
 in $C(S^1)$, for which
 the above is true, on a set of probability one, which proves the almost sure convergence stated 
 in the proposition. 

 Now assume $e^{i\lambda_0}\not\in \Sigma$ and take a continuous non negative $f$ such that
 $f(e^{i\lambda_0})=1$ and $f|_{\Sigma}=0$. Then $f(U_{\omega})=0$ a.s. so that 
 $\int f(e^{i\lambda}) dk(\lambda)=0$ and $e^{i\lambda_0}\not\in \mbox{ supp } k$. Conversely,
 if $e^{i\lambda_0}\not\in \mbox{ supp } k$, there exists a non negative continuous $f$ 
 with $f(e^{i\lambda_0})=1$ and $\int f(e^{i\lambda}) dk(\lambda)=0$. Hence, a.s., 
 $\bra \ffi_{0} |f(U_{\omega})\ffi_{0}\ket+\bra \ffi_{1} |f(U_{\omega})\ffi_{1}\ket=0$,
 therefore, by ergodicity, $\bra \ffi_{j} |f(U_{\omega})\ffi_{j}\ket=0$ a.s. for any $j$ and
 $f(U_{\omega})=0$. As $f$ is continuous and equals one at $e^{i\lambda_0}$, we get that
 $e^{i\lambda_0}\not\in \Sigma$.
 \hfill \ep

{\bf Proof of Lemma \ref{uni}:}\\
We only deal with the case where the $\theta_k^{\omega}$'s  are i.i.d. 
and uniform, the other case beeing similar. Let $\Phi_{\eta}(n)=\E(e^{in\eta_k^{\omega}})$
be the characteristic function of the random variable $\eta_k^{\omega}$, and similarly
for $\alpha_k^{\omega}$,  and  $\Phi_{\theta}(n)=\delta_{n,0}$. Then, using independence,
\be
\Phi_{\eta}(n)=\Phi_{\theta}(n)^2\Phi_{\alpha}(n)\Phi_{\alpha}(-n)=\delta_{n,0}
|\Phi_{\alpha}(n)|^2=\delta_{n,0},
\ee
so that the $\eta_k$'s are uniformly distributed. Consider now
\be
\Phi_{\eta_{k_0},\eta_{ k_1},\cdots,\eta_{ k_j}}(n_0, n_1,\cdots, n_j)=\E(e^{i\sum_{l=0}^jk_l\eta_l}).
\ee
We can assume the $k_j$'s are ordered and we observe that $\eta_k$ and $\eta_{k+j}$ are
independent as soon as $j\geq 2$, see (\ref{defeta}). Therefore, we can consider 
consecutive indices $k_l$ and deal with
\bea
 && \Phi_{\eta_{k},\eta_{ k+1},\cdots,\eta_{ k+j}}(n_1, n_2,\cdots, n_j)=\\
&&\E(e^{in_0\theta_{k-1}+i(n_0+n_1)\theta_k+\cdots+i(n_{j-1}+n_j)\theta_{k+j-1}+n_j\theta_j)}
\E(f(\alpha, \vec{n})),\nonumber
\eea
where the second expectation contains $\alpha_k$'s only. Then
\bea
&&\Phi_{\eta_{k},\eta_{ k+1},\cdots,\eta_{ k+j}}(n_1, n_2,\cdots, n_j)=\nonumber\\
&&\Phi_{\theta}(n_0)\Phi_{\theta}(n_0+n_1)\cdots\Phi_{\theta}(n_{j-1}+n_{j})
\Phi_{\theta}(n_j)\E(f(\alpha))=\nonumber\\
&&\delta_{n_0,0}\delta_{n_1,0}\cdots \delta_{n_j,0}\E(f(\alpha, \vec{n}))=
\delta_{\vec{n},\vec{0}}\E(f(\alpha, \vec{0}))=\delta_{\vec{n},\vec{0}},
\eea
whith the obvious notation, which yields the result.
\hfill\ep\\

{\bf Proof of Proposition \ref{freep}:}\\
We first prove this Proposition with the definition of the density
of states as the distribution function of the "band functions" of $U_0$,
to be defined below. Then we'll see in the course of the proof
of  Lemma \ref{deflim} below the equivalence with the definiton as an average
counting measure.
The proof of Proposition 6.2 in \cite{bhj} shows that $U_0$ on
$l^2(\Z)$ is unitarily equivalent to the operator multiplication by the matrix
\be\label{demul}
V(x)=\pmatrix{r^2-t^2e^{2ix}& 2itr\cos(x)\cr 2itr\cos(x)&r^2-t^2e^{-2ix}} \,\,\mbox{ on }
\,L^2(\T)\simeq L^2_+(\T)\oplus  L^2_-(\T),
\ee
by the unitary mapping that sends $\ffi_k\mapsto e^{ikx}/\sqrt{2\pi}$, and where 
$ L^2_{\pm}(\T)$ is the susbspace generated by even/odd harmonics $\{e^{ikx}\}_{k\in\Z}$.
The eigenvalues of $V(x)$ are 
\be\label{alpha}
  \lambda_{\pm}(x)=e^{\pm i\alpha(x)}, \,\,\mbox{ where } \,\, \alpha(x)=
\arccos(r^2-t^2\cos(2x)).
\ee
We note that $\lambda_{\pm}(x)=\lambda_{\pm}(-x)$ and 
\be
  V(x)=JV(-x)J\, \,\,\mbox{ where } \,\, J=\pmatrix{0&1\cr 1&0}.
\ee
Hence, the corresponding eigenvectors $\chi_{\pm}(x)$ satisfy
\be\label{symv}
  V(x)\chi_{\pm}(x)=\lambda_{\pm}(x)\chi_{\pm}(x) \, \,\,\mbox{ and } \, \,
V(x)J\chi_{\pm}(-x)=\lambda_{\pm}(x)J\chi_{\pm}(-x),
\ee
so that $\chi_{\pm}(x)$ and $J\chi_{\pm}(-x)$ are linearly dependent. This is in keeping with 
the fact that the subspace of generalized eigenvectors is of dimension 2, see (\ref{trans}).
Also, one checks that 
for any phase $\beta \in ]-\arccos(r^2-t^2),0[\cup ]0,\arccos(r^2-t^2)[$,
\be
\alpha^{-1}(\beta)=\{x_1,x_2,-x_2-x_1\}\subset ]-\pi, \pi[ .
\ee
Therefore, due to (\ref{symv}), only half these points contribute for the computation 
of the density of states. We can now compute the integrated density 
of states $N_0(\beta)$
as follows: Taking into account the normalisation by a factor $1/2\pi$ 
in the definition (\ref{dos}), the fact that $\mbox{ supp }k\subset[-\arccos(r^2-t^2),
\arccos(r^2-t^2)] $ and the symmetries, we have for any $\beta\in [-\arccos(r^2-t^2),0]$
\bea
N_0(\beta)&=&\frac{1}{4\pi}\int_{\T}d\lambda\chi_{\{-\alpha(\lambda )<\beta\leq 0\}}=
\frac{1}{2\pi}\int_{-\pi/2}^{\pi/2}d\lambda\chi_{\{\cos(2\lambda)>(r^2-\cos(\beta))/t^2\}}\\
&=&\frac{1}{2\pi}\int_0^{\arccos((r^2-\cos(\beta))/t^2)}=
\frac{1}{2\pi}\arccos\left(\frac{r^2-\cos(\beta)}{t^2}\right).
\eea
A similar computation for $\beta\in [0,\arccos(r^2-t^2))$ yields (\ref{nzero}).
Therefore, $dk_0$ is absolutely continuous w.r.t. Lebesgue and, for any  $|\lambda|<\arccos(r^2-t^2)$,
$dk_0(\lambda)=N'(\lambda)d\lambda$,
from which the result on the density of states follows.
In order to obtain the Lyapunov exponent, it is enough to observe that the transfer matrices (\ref{trans})
$T$, now independent of $k$, are of determinant one and trace equal to $2(r^2-\cos(\lambda))/t^2$. 
Therefore, it is readily checked that when the eigenvalues $\tau_{\pm}(\lambda)$ of $T$ 
\be
 \tau_{\pm}(\lambda)=(r^2-\cos(\lambda)\pm\sqrt{(r^2-\cos(\lambda))^2-t^4})/t^2
\ee
are complex conjugates, i.e. when $|\lambda|<\arccos(r^2-t^2)$, they are of modulus one, whereas
\be
\max\{|\tau_+|,|\tau_-|\}=(r^2-\cos(\lambda)\pm\sqrt{(r^2-\cos(\lambda))^2-t^4})/t^2 , 
\ee
if $|\lambda|\geq \arccos(r^2-t^2)$. It remains to use definition (\ref{lyapu}) to get 
$\gamma_0(e^{i\lambda})$. In order to prove the last statement, we first rewrite the right 
hand side of Thouless formula with $dk_0(\lambda')$ above as
\be\label{rew}
  \frac{1}{2\pi}\int_{-1}^1 \frac{\ln((x-y)^2)}{\sqrt{1-x^2}}dx+\ln 2
\ee
by means elementary manipulations, changing variables to $x=(r^2-\cos(\lambda'))/t^2$ and introducing
$y=(r^2-\cos(\lambda))/t^2\in [-1, (r^2+1)/t^2]$. Hence we are to show that (\ref{rew}) above equals 
$0$ if $y\leq 1$ and $\ln(y+\sqrt{y^2-1})$ if $y >1$. We first deal with the case $y>1$. We can 
differentiate (\ref{rew}) with respect to $y$ under the integral sign to get
\be\label{cint}
\frac{1}{\pi}\int_{-1}^1 \frac{dx}{\sqrt{1-x^2}(y-x)}=\frac{1}{2\pi}\int_{C}
\frac{dz}{\sqrt{1-z^2}(y-z)},
\ee
where $C$ is a contour in the complex plane surrounding the segment $[-1,1]$ in the positive 
direction which does not contain $y$ in its interior. By deforming the contour to a circle centered
at the origin and of radius $R>0$ large enough, we pick a residue at $y$. As the integral 
on the circle
is of order $1/R$, we eventually get in the limit $R\ra\infty$
\be
  \frac{d}{dy}\left\{\frac{1}{2\pi}\int_{-1}^1 \frac{\ln((x-y)^2)}{\sqrt{1-x^2}}dx+\ln 2\right \}=
\frac{1}{\sqrt{y^2-1}},
\ee
as expected. The limit as $y\ra 1^+$ of the Lyapunov exponent fixes the constant to $0$. Now, if
$y\in ]-1,1[$, we first convert (\ref{rew}) to a contour integral along a path similar to 
the one above with te following difference. As the $\ln$ is multivalued, with a cut from $y$
along the real axis towards $-\infty$, the contour is attached to the point $-1$. By assumption,
$y$ does not belong to the contour of integration, so that we can now differentiate with respect 
to $y$ under the integral sign and thus get the same contour integral (\ref{cint}) as above 
to consider. However, by expanding the contour to infinity, we get to residue this time, so that
(\ref{rew}) is constant for $y\in ]-1,1[$. As it is known (\cite{gr}, \# 4.224, p.526) that, 
\be
 \frac{1}{2\pi}\int_{-1}^1 \frac{\ln(x^2)}{\sqrt{1-x^2}}dx=\frac{2}{\pi}\int_{0}^1 
\frac{\ln(x)}{\sqrt{1-x^2}}dx=\frac{2}{\pi}\int_0^{\pi/2}\ln(\sin(t))dt=-\ln 2,
\ee
we have, by continuity, that the integral is equal to zero on $[-1,1]$.
\hfill\ep
 
{\bf Proof of Lemma \ref{deflim}:}\\
We use freely the notations above. Let us introduce the eigenprojectors
$P_{\pm}(x)$ associated with $\lambda_{\pm}(x)$ such that
\be
  V(x)=P_+(x)\lambda_+(x)+P_-(x)\lambda_-(x).
\ee
These quantities are analytic in $x$, in a strip including the real axis. 
Let $f\in C(S^1)$ and let us compute by means of (\ref{demul}) and the definition
of $L^2_{\pm}(\T)$
\bea
 &&\mbox{ tr }\bra \chi_{M,N}| f(U_0)\chi_{M,N}\ket =\sum_{M<j\leq N}\bra \ffi_j |
f(U_0)\ffi_j\ket =\nonumber\\ 
&&\sum_{ j \mbox{ \scriptsize even}\atop M<j\leq N}\frac{1}{2\pi}\int_{\T}{\Big \bra}
\pmatrix{1\cr 0}{\Big |}\left(f(\lambda_+(x))P_+(x)+f(\lambda_-(x))P_-(x)\right)
\pmatrix{1\cr 0}{\Big \ket}dx+\nonumber\\
&& \quad\quad \sum_{ j \mbox{ \scriptsize odd}\atop M<j\leq N}\frac{1}{2\pi}
\int_{\T}{\Big \bra}
\pmatrix{0\cr 1}{\Big |}\left(f(\lambda_+(x))P_+(x)+f(\lambda_-(x))P_-(x)\right)
\pmatrix{0\cr 1}{\Big \ket}dx.
\eea
The summand being independent of $j$ and uniformly bounded, we can rewrite the 
above trace as $N-M$ gets large as
\bea
&& \frac{N-M}{4\pi}\int_{\T}f(\lambda_+(x))\mbox{ tr }P_+(x)+
f(\lambda_-(x))\mbox{ tr }P_-(x)dx +O(1)=
\nonumber\\
&& \frac{N-M}{4\pi}\int_{\T}f(\lambda_+(x))+f(\lambda_-(x))dx +O(1).
\eea
Hence, with $\lambda_{\pm}(x)=e^{\pm i\alpha(x)}$ as in (\ref{alpha}),
and taking into account the properties of $\alpha$, we get
\bea
 \int_{\T}f(e^{i\lambda})dk_0(\lambda)&=& \frac{1}{4\pi}\int_{\T}
f(e^{i\alpha(x)})+f(e^{-i\alpha(x)})dx\nonumber\\
&=& \frac{1}{2\pi}\int_{-\pi/2}^{\pi/2}f(e^{i\alpha(x)})+f(e^{-i\alpha(x)})dx,
\eea
which is easily seen to coincide with the "direct" definition of $dk_0$ in the 
above proof.
\hfill\ep

{\bf Proof of Proposition \ref{better}:}\\
As in that case a commun term $\frac{1}{2^n}$ can be factorized, see (\ref{rhss}),
we compute the generating function of $|{\cal C}_{n-1}(j)|$, the cardinal of 
the set of relevant indices. Using the same symbols as above, we consider this time
\be
P_n(x)=\sum_{-2n\leq j\leq 2n}|{\cal C}_{n-1}(j)|x^j,
\ee 
 which we split into two parts 
$P_n(x)=P_n^+(x)+P_n^-(x)$ that satisfy for $n=0, 1$,
\be
P_0^+(x)=1, P_0^-(x)=0, P_1^+(x)=1+x^{-2}, P_1^-(x)=x+x^{-1}. 
\ee
As above,
\begin{lem}\label{tramat2}
For any $n\geq0$,
$$
\pmatrix{P_{n+1}^+(x)\cr P_{n+1}^-(x)}=\pmatrix{1+x^{-2} & x+x^{-1} \cr
x+x^{-1} & 1+x^{2}}\pmatrix{P_n^+(x)\cr P_n^-(x)},
$$
with $P_0^+(x)=1, P_0^-(x)=0$.
\end{lem}
By diagonalization of  the corresponding transfer matrix, we get
\be
T^n(x)=A(x)\pmatrix{0 & 0 \cr 0 & (x^{-1}+x)^{2n}}A(x)^{-1}
\ee
where 
\be
A(x)=\pmatrix{1+x^{2} & x+x^{-1} \cr
-(x+x^{-1}) & 1+x^{2}}
\ee
and we compute
\be
\pmatrix{P_n^+(x)\cr P_n^-(x)}=T^n(x)\pmatrix{1\cr 0}=\frac{(x^2+1)^{2n-1}}{x^{2n}}
\pmatrix{1\cr x}.
\ee
Using the binomial Theorem we obtain for $P_n^{\pm}(x)$
\bea
P_n^+(x)&=&\sum_{l=-n}^{n-1}x^{2l}\pmatrix{2n-1 \cr l+n}\nonumber\\
P_n^-(x)&=&\sum_{l=-n}^{n-1}x^{2l+1}\pmatrix{2n-1 \cr l+n},
\eea
hence the end result. \hfill \ep\\

{\bf Acknowledgements:}\\
It is a pleasure to thank O.Bourget and R.Bacher for useful discussions and
D.Damanik for pointing out reference \cite{cmv} to me.

\end{document}